\pdfoutput=1
\documentclass[12pt,a4paper]{article}

\usepackage{ifthen} 
\newboolean{pdflatex}
\setboolean{pdflatex}{true} 

\newboolean{articletitles}
\setboolean{articletitles}{true} 

\newboolean{uprightparticles}
\setboolean{uprightparticles}{false} 

\def\paperauthors{LHCb collaboration} 
\def\paperasciititle{Search for K0->pipimumu decays at LHCb} 
\def\papertitle{Search for \Kzppmm decays at LHCb} 
\def\paperkeywords{{High Energy Physics}, {LHCb}} 
\def\papercopyright{\the\year\ CERN for the benefit of the LHCb collaboration} 
\def\paperlicence{CC BY 4.0 licence}
\def\paperlicenceurl{https://creativecommons.org/licenses/by/4.0/}

\newif\ifEnableSectionTOCLinks
\EnableSectionTOCLinksfalse 

\usepackage[top=1in, bottom=1.25in, left=1in, right=1in]{geometry}

\usepackage{microtype}
\usepackage{lineno}  
\usepackage{xspace} 
\usepackage{caption} 

\usepackage{booktabs}

\usepackage{graphicx}  
\usepackage{color}
\usepackage{colortbl}
\graphicspath{{./figs/}} 

\usepackage{amsmath} 
\usepackage{amssymb}
\usepackage{amsfonts}
\usepackage{upgreek} 

\newcommand*\patchAmsMathEnvironmentForLineno[1]{%
\expandafter\let\csname old#1\expandafter\endcsname\csname #1\endcsname
\expandafter\let\csname oldend#1\expandafter\endcsname\csname
end#1\endcsname
 \renewenvironment{#1}%
   {\linenomath\csname old#1\endcsname}%
   {\csname oldend#1\endcsname\endlinenomath}%
}
\newcommand*\patchBothAmsMathEnvironmentsForLineno[1]{%
  \patchAmsMathEnvironmentForLineno{#1}%
  \patchAmsMathEnvironmentForLineno{#1*}%
}
\AtBeginDocument{%
\patchBothAmsMathEnvironmentsForLineno{equation}%
\patchBothAmsMathEnvironmentsForLineno{align}%
\patchBothAmsMathEnvironmentsForLineno{flalign}%
\patchBothAmsMathEnvironmentsForLineno{alignat}%
\patchBothAmsMathEnvironmentsForLineno{gather}%
\patchBothAmsMathEnvironmentsForLineno{multline}%
\patchBothAmsMathEnvironmentsForLineno{eqnarray}%
}

\usepackage[pdftex,
            pdfauthor={\paperauthors},
            pdftitle={\paperasciititle},
            pdfkeywords={\paperkeywords}]{hyperref}
\usepackage{hyperxmp}
\hypersetup{
    pdfcopyright={Copyright (C) \papercopyright},
    pdflicenseurl={\paperlicenceurl}
}

\usepackage[colorinlistoftodos,textsize=scriptsize]{todonotes}

\usepackage[bottom,flushmargin,hang,multiple]{footmisc}

\usepackage[all]{hypcap} 

\usepackage{xspace}
\usepackage{upgreek}

\def\lhcb   {\mbox{LHCb}\xspace}

\def\lhc    {\mbox{LHC}\xspace}

\def\MagUp {\mbox{\em Mag\kern -0.05em Up}\xspace}

\def\lone   {L0\xspace}
\def\hlt    {HLT\xspace}
\def\hltone {HLT1\xspace}
\def\hlttwo {HLT2\xspace}

\ifthenelse{\boolean{uprightparticles}}%
{
 
 \def\Pgamma      {\ensuremath{\upgamma}\xspace}

 \def\Pmu         {\ensuremath{\upmu}\xspace}

 \def\Ppi         {\ensuremath{\uppi}\xspace}

 \def\Ppsi        {\ensuremath{\uppsi}\xspace}

 \def\PDelta      {\ensuremath{\Delta}\xspace}
 \def\PXi         {\ensuremath{\Xi}\xspace}
 \def\PLambda     {\ensuremath{\Lambda}\xspace}
 \def\PSigma      {\ensuremath{\Sigma}\xspace}
 \def\POmega      {\ensuremath{\Omega}\xspace}
 \def\PUpsilon    {\ensuremath{\Upsilon}\xspace}
 \let\oldPi\Pi
 \def\PPi         {\ensuremath{\oldPi}\xspace}

 \def\PB      {\ensuremath{\mathrm{B}}\xspace}
 \def\PD      {\ensuremath{\mathrm{D}}\xspace}
 \def\PJ      {\ensuremath{\mathrm{J}}\xspace}
 \def\PK      {\ensuremath{\mathrm{K}}\xspace}
 \def\Pb      {\ensuremath{\mathrm{b}}\xspace}
 \def\Pc      {\ensuremath{\mathrm{c}}\xspace}
 
 \def\Pe      {\ensuremath{\mathrm{e}}\xspace}
 \def\Pp      {\ensuremath{\mathrm{p}}\xspace}

 \def\Ps      {\ensuremath{\mathrm{s}}\xspace}

 \def\thebaroffset{0.0em}
}
{
 
 \def\Pgamma      {\ensuremath{\gamma}\xspace}

 \def\Pmu         {\ensuremath{\mu}\xspace}

 \def\Ppi         {\ensuremath{\pi}\xspace}

 \def\Ppsi        {\ensuremath{\psi}\xspace}
 
 \mathchardef\PDelta="7101
 \mathchardef\PXi="7104
 \mathchardef\PLambda="7103
 \mathchardef\PSigma="7106
 \mathchardef\POmega="710A
 \mathchardef\PUpsilon="7107
 \mathchardef\PPi="7105
 \def\PB      {\ensuremath{B}\xspace}
 \def\PD      {\ensuremath{D}\xspace}
 \def\PJ      {\ensuremath{J}\xspace}
 \def\PK      {\ensuremath{K}\xspace}
 \def\Pb      {\ensuremath{b}\xspace}
 \def\Pc      {\ensuremath{c}\xspace}
 
 \def\Pe      {\ensuremath{e}\xspace}
 \def\Pp      {\ensuremath{p}\xspace}

 \def\Ps      {\ensuremath{s}\xspace}

 \def\thebaroffset{0.18em}
}
\newcommand{\offsetoverline}[2][\thebaroffset]{\kern #1\overline{\kern -#1 #2}}%

\makeatletter
\ifcase \@ptsize \relax
  \newcommand{\miniscule}{\@setfontsize\miniscule{4}{5}}
\or
  \newcommand{\miniscule}{\@setfontsize\miniscule{5}{6}}
\or
  \newcommand{\miniscule}{\@setfontsize\miniscule{5}{6}}
\fi
\makeatother

\DeclareRobustCommand{\optbar}[1]{\shortstack{{\miniscule (\rule[.5ex]{1.25em}{.18mm})}
  \\ [-.7ex] $#1$}}

\def\electron   {{\ensuremath{\Pe}}\xspace}
\def\en         {{\ensuremath{\Pe^-}}\xspace}   
\def\ep         {{\ensuremath{\Pe^+}}\xspace}

\def\muon       {{\ensuremath{\Pmu}}\xspace}

\def\mump       {{\ensuremath{\Pmu^\mp}}\xspace}
\def\mumu       {{\ensuremath{\Pmu^+\Pmu^-}}\xspace}

\def\ellm       {{\ensuremath{\ell^-}}\xspace}
\def\ellp       {{\ensuremath{\ell^+}}\xspace}

\def\ellell     {\ensuremath{\ellp \ellm}\xspace}

\def\g      {{\ensuremath{\Pgamma}}\xspace}

\def\squark    {{\ensuremath{\Ps}}\xspace}

\def\cquark    {{\ensuremath{\Pc}}\xspace}

\def\bquark    {{\ensuremath{\Pb}}\xspace}

\def\pion   {{\ensuremath{\Ppi}}\xspace}

\def\pip    {{\ensuremath{\pion^+}}\xspace}
\def\pim    {{\ensuremath{\pion^-}}\xspace}
\def\pipm   {{\ensuremath{\pion^\pm}}\xspace}

\def\kaon    {{\ensuremath{\PK}}\xspace}
\def\Kbar    {{\ensuremath{\offsetoverline{\PK}}}\xspace}

\def\KorKbar {\kern \thebaroffset\optbar{\kern -\thebaroffset \PK}{}\xspace}
\def\Kz      {{\ensuremath{\kaon^0}}\xspace}
\def\Kzb     {{\ensuremath{\Kbar{}^0}}\xspace}

\def\KS      {{\ensuremath{\kaon^0_{\mathrm{S}}}}\xspace}

\def\KL      {{\ensuremath{\kaon^0_{\mathrm{L}}}}\xspace}

\def\D       {{\ensuremath{\PD}}\xspace}

\def\DorDbar {\kern \thebaroffset\optbar{\kern -\thebaroffset \PD}\xspace}

\def\Dp      {{\ensuremath{\D^+}}\xspace}
\def\Dm      {{\ensuremath{\D^-}}\xspace}

\def\DpDm    {\ensuremath{\Dp {\kern -0.16em \Dm}}\xspace}

\def\B       {{\ensuremath{\PB}}\xspace}

\def\BorBbar {\kern \thebaroffset\optbar{\kern -\thebaroffset \PB}\xspace}

\def\Bd      {{\ensuremath{\B^0}}\xspace}

\def\BdorBdbar {\kern \thebaroffset\optbar{\kern -\thebaroffset \Bd}\xspace}

\def\Bs      {{\ensuremath{\B^0_\squark}}\xspace}

\def\BsorBsbar {\kern \thebaroffset\optbar{\kern -\thebaroffset \Bs}\xspace}

\def\jpsi     {{\ensuremath{{\PJ\mskip -3mu/\mskip -2mu\Ppsi}}}\xspace}

\def\Y#1S{\ensuremath{\PUpsilon{(#1S)}}\xspace}

\def\proton      {{\ensuremath{\Pp}}\xspace}
\def\antiproton  {{\ensuremath{\overline \proton}}\xspace}

\def\Lz          {{\ensuremath{\PLambda}}\xspace}
\def\Lbar        {{\ensuremath{\offsetoverline{\PLambda}}}\xspace}
\def\LorLbar     {\kern \thebaroffset\optbar{\kern -\thebaroffset \PLambda}\xspace}

\def\BF         {{\ensuremath{\mathcal{B}}}\xspace}

\newcommand{\decay}[2]{\ensuremath{\mathinner{#1\!\to #2}}\xspace}

\def\to                 {\ensuremath{\rightarrow}\xspace}

\def\CP                {{\ensuremath{C\!P}}\xspace}

\def\AT#1     {\ensuremath{A_{\mathrm{T}}^{#1}}\xspace}           

\def\C#1      {\ensuremath{\mathcal{C}_{#1}}\xspace}                       
\def\Cp#1     {\ensuremath{\mathcal{C}_{#1}^{'}}\xspace}                    
\def\Ceff#1   {\ensuremath{\mathcal{C}_{#1}^{\mathrm{(eff)}}}\xspace}        
\def\Cpeff#1  {\ensuremath{\mathcal{C}_{#1}^{'\mathrm{(eff)}}}\xspace}       
\def\Ope#1    {\ensuremath{\mathcal{O}_{#1}}\xspace}                       
\def\Opep#1   {\ensuremath{\mathcal{O}_{#1}^{'}}\xspace}                    

\newcommand{\nospaceunit}[1]{\ensuremath{\text{#1}}}
\newcommand{\aunit}[1]{\ensuremath{\text{\,#1}}}

\newcommand{\tev}{\aunit{Te\kern -0.1em V}\xspace}
\newcommand{\gev}{\aunit{Ge\kern -0.1em V}\xspace}
\newcommand{\mev}{\aunit{Me\kern -0.1em V}\xspace}
\newcommand{\kev}{\aunit{ke\kern -0.1em V}\xspace}
\newcommand{\ev}{\aunit{e\kern -0.1em V}\xspace}

\newcommand{\mevc}{\ensuremath{\aunit{Me\kern -0.1em V\!/}c}\xspace}
\newcommand{\gevc}{\ensuremath{\aunit{Ge\kern -0.1em V\!/}c}\xspace}
\newcommand{\mevcc}{\ensuremath{\aunit{Me\kern -0.1em V\!/}c^2}\xspace}
\newcommand{\gevcc}{\ensuremath{\aunit{Ge\kern -0.1em V\!/}c^2}\xspace}

\def\mum  {\ensuremath{\,\upmu\nospaceunit{m}}\xspace}

\def\barn{\aunit{b}\xspace}

\def\fb   {\ensuremath{\aunit{fb}}\xspace}
\def\invfb   {\ensuremath{\fb^{-1}}\xspace}

\newcommand{\chisq}{\ensuremath{\chi^2}\xspace}

\def\gsim{{~\raise.15em\hbox{$>$}\kern-.85em
          \lower.35em\hbox{$\sim$}~}\xspace}
\def\lsim{{~\raise.15em\hbox{$<$}\kern-.85em
          \lower.35em\hbox{$\sim$}~}\xspace}

\def\pt         {\ensuremath{p_{\mathrm{T}}}\xspace}

\def\ptot       {\ensuremath{p}\xspace}

\def\evtgen     {\mbox{\textsc{EvtGen}}\xspace}

\def\geant      {\mbox{\textsc{Geant4}}\xspace}

\def\photos     {\mbox{\textsc{Photos}}\xspace}

\def\pythia     {\mbox{\textsc{Pythia}}\xspace}

\def\tell1  {TELL1\xspace}
\def\ukl1   {UKL1\xspace}

\newcommand{\lhcborcid}[1]{\href{https://orcid.org/#1}{\hspace*{0.1em}\raisebox{-0.45ex}{\includegraphics[width=1em]{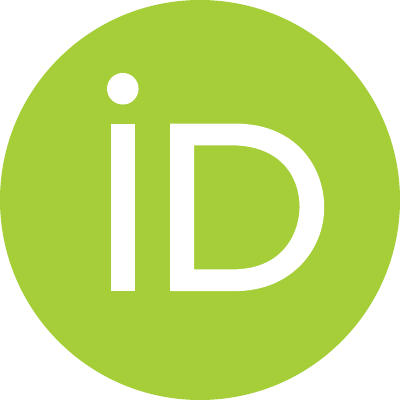}}}}

\hypersetup{
  colorlinks   = true, 
  urlcolor     = blue, 
  linkcolor    = blue, 
  citecolor    = red   
}

\ifEnableSectionTOCLinks
    \usepackage[explicit]{titlesec} 
    
    \let\oldcontentsline\contentsline
    \renewcommand

    \titleformat{\section}{\normalfont\Large\bf}{\hyperlink{tocsection.\thesection}{{\thesection} \parbox[t]{\dimexpr\textwidth-1pc}{#1}}}{1pc}{}

    \titleformat{\subsection}{\normalfont\bf}{\hyperlink{tocsubsection.\thesubsection}{{\thesubsection} \parbox[t]{\dimexpr\textwidth-1pc}{#1}}}{1pc}{}

    \titleformat{name=\section,numberless}[display]{}{}{0pt}{\normalfont\Huge\bfseries #1}
\fi

\usepackage{cite} 
\usepackage{mciteplus}

\newcommand{\pipi}{\ensuremath{\Ppi^+\Ppi^-}\xspace}

\newcommand{\Kl}{\KL}

\newcommand{\Ksmumu}{\decay{\KS}{\mumu}}

\newcommand{\Kspipi}{\decay{\KS}{\pip\pim}}

\newcommand{\KzFM}{\ensuremath{K^0_{\textrm{S(L)}}\to\mu^+\mu^-\mu^+\mu^-}\xspace}
\newcommand{\Kppmm}{\decay{\KS}{\pipi\mumu}}
\newcommand{\Klppmm}{\decay{\KL}{\pipi\mumu}}
\newcommand{\Kzppmm}{\decay{K^0_{\textrm{S(L)}}}{\pipi\mumu}}

\newcommand{\Lambdappi}{\decay{\Lz}{\proton\pim}}
\newcommand{\Lambdabppi}{\decay{\Lbar}{\antiproton\pip}}

\newcommand{\BRof}[1]{\ensuremath{{\cal B}(#1)}\xspace}

\newcommand{\IP}{\ensuremath{{\rm IP}}\xspace}

\newcommand{\figref}[1]{Fig.~\ref{#1}}

\newcommand{\tabref}[1]{Table~\ref{#1}}

\def\nuornubar {\kern \thebaroffset\optbar{\kern -\thebaroffset \nu}\xspace}
\usepackage{tikz}
\usepackage{tikz-feynman}

\begin{document}

\renewcommand{\thefootnote}{\fnsymbol{footnote}}
\setcounter{footnote}{1}


\begin{titlepage}
\pagenumbering{roman}

\vspace*{-1.5cm}
\centerline{\large EUROPEAN ORGANIZATION FOR NUCLEAR RESEARCH (CERN)}
\vspace*{1.5cm}
\noindent
\begin{tabular*}{\linewidth}{lc@{\extracolsep{\fill}}r@{\extracolsep{0pt}}}
\vspace*{-1.5cm}\mbox{\!\!\!\includegraphics[width=.14\textwidth]{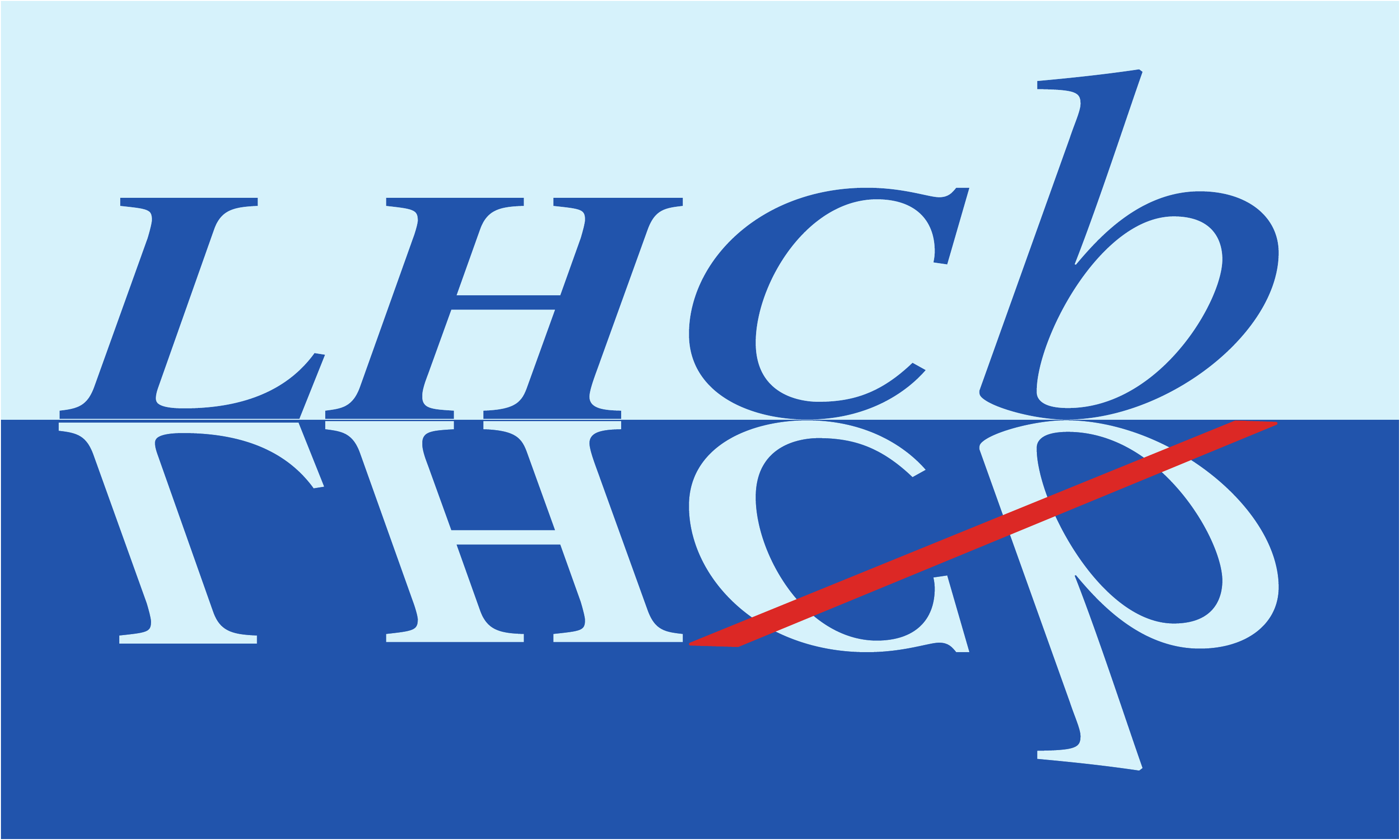}} & &%
\\
 & & CERN-EP-2025-227 \\  
 & & LHCb-PAPER-2025-045 \\  
 & & May 6, 2026 \\ 
 & & \\
\end{tabular*}

\vspace*{4.0cm}

{\normalfont\bfseries\boldmath\huge
\begin{center}
  \papertitle 
\end{center}
}

\vspace*{2.0cm}

\begin{center}
\paperauthors\footnote{Authors are listed at the end of this paper.}
\end{center}

\vspace{\fill}

\begin{abstract}
  \noindent
  A search for $K_{\textrm{S(L)}}^0\rightarrow\pi^{+}\pi^{-}\mu^{+}\mu^{-}$ decays is performed 
  using proton-proton collision data collected by the LHCb experiment at a centre-of-mass energy of 13\tev, 
  corresponding to an integrated luminosity of 5.4\invfb. No $K_{\textrm{S(L)}}^0\rightarrow\pi^{+}\pi^{-}\mu^{+}\mu^{-}$ signals are found and upper limits are set for the first time on the branching fractions 
  \mbox{\BRof\Kppmm $< 1.4\times 10^{-9}$} and \mbox{\BRof\Klppmm $< 6.6\times 10^{-7}$,} at the 90\% confidence level.
  
\end{abstract}

\vspace*{2.0cm}

\begin{center}
  Published in Phys. Rev. D 113 (2026) 072001
\end{center}

\vspace{\fill}

{\footnotesize 
\centerline{\copyright~\papercopyright. \href{\paperlicenceurl}{\paperlicence}.}}
\vspace*{2mm}

\end{titlepage}


\newpage
\setcounter{page}{2}
\mbox{~}
%
%
%
%


\renewcommand{\thefootnote}{\arabic{footnote}}
\setcounter{footnote}{0}


\cleardoublepage


\pagestyle{plain} 
\setcounter{page}{1}
\pagenumbering{arabic}


\section{Introduction}
\label{sec:introduction}

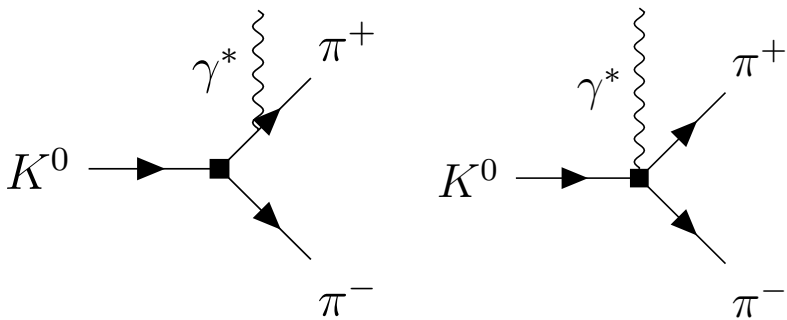
\begin{figure}[b]
    \centering
    \resizebox{0.35\textwidth}{!}{
    \begin{tikzpicture}
    \begin{feynman}
        \vertex (p) at (-10, 0) {\Kz};
        \vertex [right=of p] (v1) [square dot] {};
        \vertex [above right=of v1] (p1) {\pip};
        \vertex [below right=of v1] (p2) {\pim};
        \vertex [above=of p1] (g1) ; 
        
    \coordinate (mid) at ($(v1)!0.3!(p1)$);
    \coordinate (g1) at ($(mid)+(0,1)$);
        
        \diagram* {
            (p) -- [fermion] (v1),
            (v1) -- [fermion, label=\pip] (p1),
            (v1) -- [fermion, label=\pim] (p2),
            (mid) -- [photon, edge label=\(\gamma^\ast\)] (g1),
                 
        };
    \end{feynman}
\end{tikzpicture}}
    \resizebox{0.33\textwidth}{!}{
    \begin{tikzpicture}
    \begin{feynman}
        \vertex (p) at (-10, 0) {\Kz};
        \vertex [right=of p] (v1) [square dot] {};
        \vertex [above right=of v1] (p1) {\pip};
        \vertex [below right=of v1] (p2) {\pim};
        \vertex [above=of v1] (g1) ; 
        
        \diagram* {
            (p) -- [fermion] (v1),
            (v1) -- [fermion, label=\pip] (p1),
            (v1) -- [fermion, label=\pim] (p2),
            (v1) -- [photon, edge label=\(\gamma^\ast\)] (g1),
                 
        };
    \end{feynman}
\end{tikzpicture}}
    \caption{Tree-level diagrams in chiral perturbation theory for the \decay{\Kz}{\pipi\g^\ast} decay with (left) inner bremsstrahlung and (right) direct emission of a $\g^\ast$ photon.}
    \label{fig:pipig_dia}
\end{figure}

Radiative decays of neutral kaons, in which a virtual photon converts into a Dalitz pair of leptons, provide a powerful laboratory for testing chiral perturbation theories and studying \CP violation~\cite{Cappiello:2017ilv}. In particular, the \decay{\Kz}{\pipi\g^\ast}, with \decay{\g^\ast}{\ell\ell}, decay proceeds through two different processes: inner bremsstrahlung, in which the virtual photon is emitted from the external charged pions; and direct emission, where the photon is generated at the weak vertex (Fig.~\ref{fig:pipig_dia}). 
The \CP information is encoded in the photon polarisation and, equivalently, in the distribution of the angle between the dipion and dilepton planes~\cite{Sehgal:1992wm}. As the \Kl and \KS mesons have opposite \CP eigenvalues, the \decay{\Kl}{\pipi\g^\ast} and \decay{\KS}{\pipi\g^\ast} decays provide complementary constraints. The short-lived mode is dominated by the \decay{\KS}{\pipi} process with inner bremsstrahlung and gives access to the long-distance (LD) chiral terms, while the long-lived mode is dominated by the direct emission (as the \decay{\KL}{\pipi} decay is heavily suppressed) and grants access to \CP-violating terms~\cite{Sehgal:1992wm}.

Dalitz decays of the electron modes, $K_{\textrm{S(L)}}^0\rightarrow\pip\pim \ep\en$, have been observed by the KTeV~\cite{KTeV:1998xwb} and KEK~\cite{Takeuchi:1998vr} collaborations. The most recent and precise results are provided by the  NA48 collaboration. Their branching fractions 
$\BF(\KL\to\pip\pim \ep \en) = (3.08\pm 0.09 \pm 0.18)\times10^{-7}$, and
$\BF(\KS\to\pip\pim \ep\en) = (4.83\pm 0.11\pm0.14)\times10^{-5}$~\cite{NA48:2003pwz,NA481:2011gxv},
differ by two orders of magnitude, as expected due to the opposite \CP eigenvalues~\cite{Sehgal:1992wm,Heiliger:1993qt}.

The equivalent muonic mode, \Kzppmm, probes higher values of lepton-pair invariant mass ($q$), and thus has larger sensitivity to \CP-violating observables where the direct emission contribution is enhanced with respect to the electron mode. 
The \Kzppmm decay is highly suppressed in the Standard Model  due to the phase-space limitations
presented by the $\sim7\mevcc$ difference in mass between the parent $K_{\textrm{S(L)}}^0$ and the four final-state particles.
Theoretical calculations of $\BF(\Kppmm)$  are dominated by LD contributions 
to the $\KS\to\pipi\gamma^\ast$ decay, and give 
    $\BRof\Kppmm = (417.00 + 2.17 + 49.80)\times10^{-16}$~\cite{Cappiello:2017ilv}, 
where the first term is due to a bremsstrahlung photon emitted by the pion, 
the second term accounts for the direct emission of the photon at the weak vertex,
and the last term represents their interference. 
Theoretical calculations for $\BF(\Kl\to\pip\pim\ellell)$ have been performed in Ref.~\cite{Sehgal:1992wm,Pichl:2000ab,Heiliger:1993qt} for $\ell=\electron$, but no specific prediction is given for  \BRof\Klppmm.
Experimentally, neither the \Kppmm nor the \Klppmm decays have been searched for before. 

The \lhcb experiment~\cite{Junior:2018odx} has proven its strong capabilities to study rare kaon decays thanks to its excellent tracking and particle-identification (PID) performance, versatile trigger system and large hadron production rate.
The \lhcb collaboration has previously set world-best upper limits on $\BF(\Ksmumu)$~\cite{LHCb-PAPER-2019-038} and $\BF(\KzFM)$~\cite{LHCB-PAPER-2022-035}.

This paper presents the first-ever search for \Kzppmm, 
using proton-proton ($pp$) collision data collected by the \lhcb experiment
during 2016--2018 at a centre-of-mass energy of $13\tev$, and 
corresponding to an integrated luminosity of  $5.4\invfb$. 

\section{Analysis strategy}

The branching fractions are measured with respect to the \Kspipi decay, according to  
\begin{equation}
    \BRof\Kzppmm = \alpha_{\textrm{S(L)}} N_{\pipi\mumu},
\end{equation}
\noindent where the normalisation factor $\alpha_{\textrm{S(L)}}$, also known as single-event sensitivity,\footnote{The value of \BRof\Kppmm corresponding to one signal event seen in the data sample.}  is defined as
\begin{equation}
    \alpha_{\textrm{S(L)}} \equiv \BRof\Kspipi\frac{\varepsilon_{\KS \to \pipi}}{N_{\pipi}\,{\varepsilon_{K^0_{\textrm{S(L)}} \to \pipi\mumu}}}.
    \label{eq:alpha}
\end{equation}
\noindent The parameter $\varepsilon_{\Kspipi} (\varepsilon_{K^0_{\textrm{S(L)}}\to\pipi\mumu})$ represents the product of the reconstruction, trigger and selection 
efficiency of \Kspipi (\Kzppmm) candidates. The observed number of $K^0_{\textrm{S(L)}}\to\pipi\mumu$ candidates is $N_{\pipi\mumu}$, the observed number of \Kspipi candidates is $N_{\pipi}$, and $\BRof\Kspipi=(69.20\pm0.05)\%$~\cite{PDG2024} is the branching fraction of the normalisation channel. The \Kspipi normalisation channel is selected due to its abundant rate. Furthermore, it enables cancellation of the \Kz production rate and reduces systematic uncertainties related to final-state particle reconstruction.

At \lhcb, the \Klppmm and \Kppmm decays are indistinguishable on an event-by-event basis~\cite{Junior:2018odx}. 
For this reason, the \KL contribution is not considered when setting the upper limit for \Kppmm, and vice versa. As a consequence, the set upper limits are more conservative than they would be if the \KS and \KL modes could be distinguished.
The \lhcb acceptance for \KL decays is $\sim2\times10^{-3}$ times that of the \KS modes~\cite{Junior:2018odx}. The only relevant difference between \Kppmm and \Klppmm at reconstruction level is the lifetime. As a consequence, the selection strategy is optimized for \Kppmm decays and then applied to \Klppmm decays. 
The \KS--\KL interference is ignored as \Kz and \Kzb mesons are produced in similar amounts at \lhc~\cite{DAmbrosio:2025mxa}.

\section{\lhcb detector and trigger}
\label{sec:lhcb}

The \lhcb detector~\cite{LHCb-DP-2008-001,LHCb-DP-2014-002} is a single-arm forward spectrometer covering the pseudorapidity range 
$2 < \eta < 5$, designed for the study of particles containing \bquark or \cquark quarks.
The detector used for this analysis includes a high-precision tracking system consisting of a silicon-strip vertex detector (VELO) surrounding the $pp$ interaction region~\cite{LHCb-DP-2014-001}, a large-area silicon-strip detector located upstream 
of a dipole magnet with a bending power of about $4{\mathrm{\,T\,m}}$, and three stations of silicon-strip detectors 
and straw drift tubes~\cite{LHCb-DP-2017-001} placed downstream of the magnet. The tracking system provides a measurement of the momentum (\ptot), of charged particles with a relative uncertainty that varies from 0.5\% at low momentum to 1.0\% at 200\gevc. The minimum distance of a track to a \textit{pp} collision vertex (PV),
the impact parameter (IP), is measured with a resolution of $(15+29/\pt)\mum$, where \pt is the component of the momentum transverse to the beam axis, in \gevc. Different types
of charged hadrons are distinguished using information from two ring-imaging Cherenkov
(RICH) detectors~\cite{LHCb-DP-2012-003}. Photons, electrons and hadrons are identified by a calorimeter system consisting of scintillating-pad and preshower detectors, an electromagnetic and a hadronic calorimeter. Muons are identified by a system composed of alternating layers of iron
and multiwire proportional chambers~\cite{LHCb-DP-2012-002}. In addition, information from the tracking system, the calorimeter system and the RICH detectors is used to further improve the muon
identification. 

Events are first filtered by a hardware trigger, known as \lone~\cite{LHCb-DP-2012-004}, which uses information from the calorimeter and the muon system to select muon candidates with \pt above a few \gevc. 
Subsequently, a full event reconstruction is applied in the High Level Trigger (HLT), which is subdivided into two steps, \hltone and \hlttwo. 
The \hltone algorithms for strange decays were improved in 2016~\cite{Aaij:2253050,Dettori:2297352}, boosting the efficiency of the \Kppmm channel, amongst others. For this reason, only the 2016--2018 data is considered in this measurement. 
In the offline selection, trigger signals are associated with reconstructed particles. Selection requirements can therefore be made on the trigger selection itself and on whether the decision was due to the signal candidate (trigger on signal, TOS) or other particles produced in the \textit{pp} collision (trigger independent of signal, TIS). To enhance the \lone trigger efficiency, events fulfilling either the TIS or TOS requirements are considered in this analysis. 
The \Kspipi normalisation channel candidates are reconstructed from data selected by a trigger line requiring only the presence of a $pp$ interaction, providing an unbiased representation of the collision. 
The triggered data further undergo a centralised, offline processing step~\cite{Stripping}.

Simulation is used to model the effects of the detector acceptance and
selection requirements. 
In the simulation, $pp$ collisions are generated using \pythia8~\cite{Sjostrand:2007gs,Sjostrand:2006za} 
  with a specific \lhcb configuration~\cite{LHCb-PROC-2010-056}.
  Decays of unstable particles
  are described by \evtgen~\cite{Lange:2001uf}, in which final-state
  radiation is generated using \photos~\cite{davidson2015photos}.
  The interaction of the generated particles with the detector, and its response,
  are implemented using the \geant
  toolkit~\cite{Allison:2006ve, Agostinelli:2002hh} as described in
  Ref.~\cite{LHCb-PROC-2011-006}. 
  The simulated samples are corrected to account for known data-simulation differences in the \KS production kinematics, as well as the track reconstruction and  PID efficiencies.
  The analysis strategy is developed using \Kppmm simulated samples. Given the \KL decay only differs due to its lifetime, the \Klppmm efficiency is obtained by weighting the decay time distribution of the simulated \Kppmm sample to match that of the \KL meson. 
  
\section{Signal selection}
\label{sec:selection}

This measurement benefits from the large kaon production cross-section at the LHC, approximately $0.3\barn$ for both \KS and \KL mesons~\cite{Junior:2018odx}, as well as from the forward production of kaons, which fall within the \lhcb detector acceptance.
The \KS candidates from both the signal and the normalisation channels are required to originate from a PV. In case of multiple choices, the PV that fits best to the flight direction of the \KS candidate is chosen as the associated PV.
Candidate \Kppmm decays are reconstructed by combining pairs 
of oppositely charged well-identified muons and pions originating from a common vertex, displaced from the PV. 
Candidates with an invariant mass in the region $490 < m(\pipi\mumu) < 600\mevcc$ are retained for further analysis. 
In order to avoid experimenter's bias, 
selected candidates in the region $490 < m({\pipi\mumu}) < 510\mevcc$, which contains approximately $97\%$ of the signal, were removed from the data sample until the analysis procedure had been finalised.

Given the relatively long lifetime of the \KS meson, only about a third of its decays occur inside the VELO, while the remainder take place downstream in the detector. Decays inside the VELO can be reconstructed with an invariant-mass resolution roughly twice as good as those outside. Moreover, the online event selection relies on tracks with hits in both the VELO and the downstream tracking stations. Consequently, only decays taking place inside the VELO are retained in this analysis.

Several selections are applied to reconstructed candidates in order to further reduce background contributions. 
These include requirements on the maximum distance of closest approach between any pair of the four signal tracks,
the flight distance from the PV of the \KS candidate on the plane transverse to the beam direction,
the angles between the two muon and the two pion tracks,
and PID variables of the final-state pions and muons.

The only non-negligible background source in the signal channel arises 
from random combinations of tracks, which originate from a $pp$ interaction region 
or from inelastic interactions of the products of the \proton\proton collision with the detector material. 
A boosted decision tree~(BDT)~\cite{Breiman,AdaBoost} classifier implemented in the 
\textsc{XGBoost} package~\cite{XGBoost} is used 
to further separate signal from combinatorial background. 
One BDT per trigger category (TIS and TOS) is used, each including as discriminating variables:

\begin{itemize}
  \item \label{var:vmv} the significance of the distance between the candidate decay vertex and any element of the VELO material~\cite{LHCb-DP-2018-002};
  \item the smallest \IP value of each of the muon and pion candidates with respect to any of the PVs reconstructed in the event;
  \item the smallest \IP value of the \KS candidate to any of the PVs reconstructed in the event;
  \item the difference in \chisq of the PV fit when the \KS candidate is or is not included;
  \item the transverse displacement of the \KS decay vertex from the beam line;
  \item the maximum distance of closest approach between any pair of the four final-state tracks.
\end{itemize}
The BDT is trained with simulated \Kppmm decays as a proxy for the signal, and \pipi\mumu combinations with an invariant mass $m(\pipi\mumu)>510\mevcc$ as a proxy for background.
To mitigate the effects of statistical fluctuations and minimise overtraining, the $k$-fold method~\cite{Scikit-learn-paper}, with $k=6$, is used during the training. 

The BDT requirement is optimized for the best expected upper limit at the 90\% confidence level~(CL) using pseudoexperiments, performing fits as explained below. 
The BDT requirement corresponds to a signal efficiency of approximately $80\%$ (95\%) for the TIS (TOS) sample, and a background suppression above 99.9\%. 
After the BDT selection requirement, in the case of multiple candidates per event, a single candidate is chosen arbitrarily.

\section{Fit model and normalisation decay mode}
\label{sec:normalisation}

\begin{figure}[tb]
    \centering
    \includegraphics[width=0.6\textwidth]{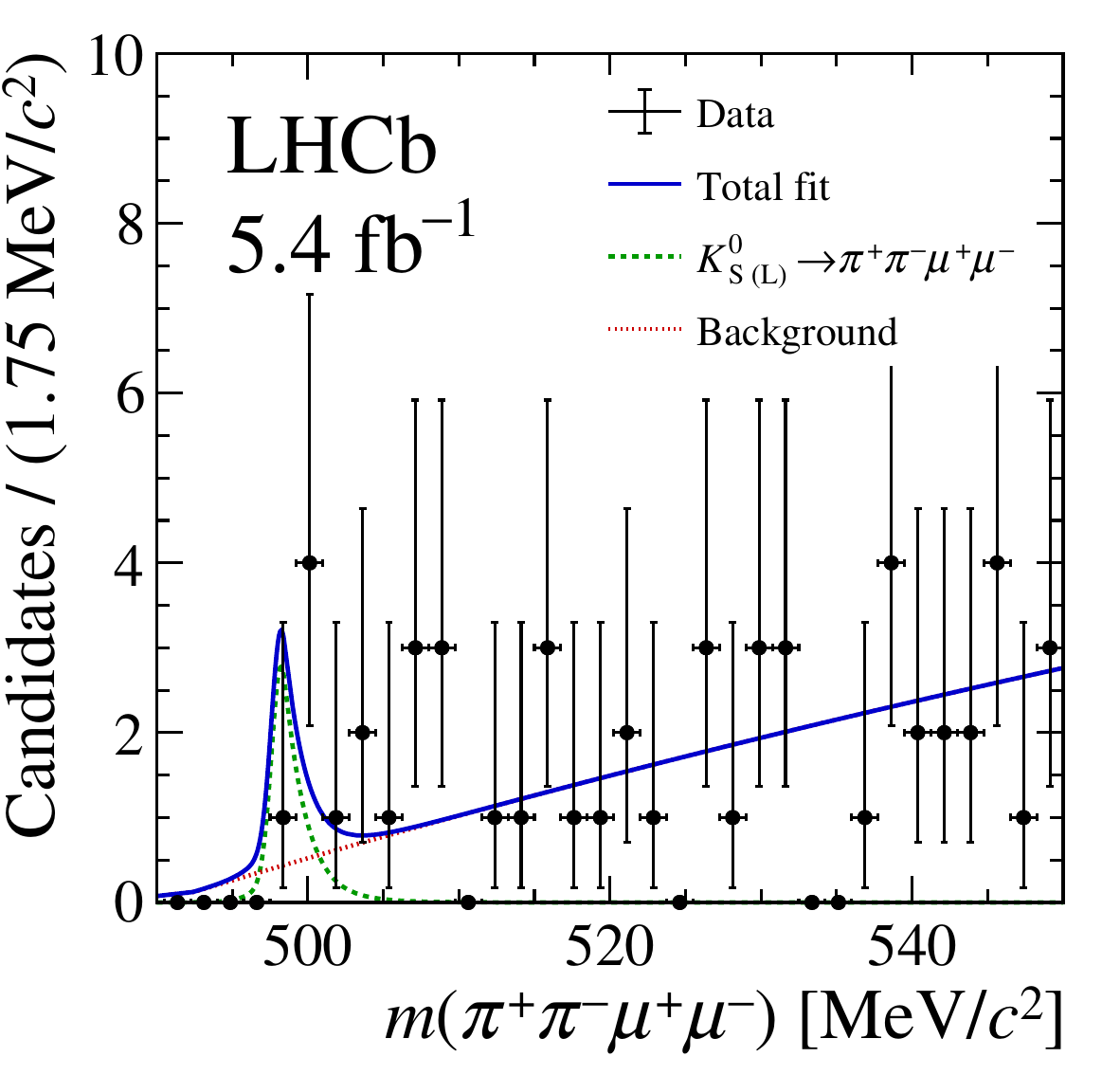}
    \caption{Invariant-mass distribution of selected \Kzppmm candidates in the TIS and TOS trigger categories combined. The blue lines show the result of a simultaneous fit to both categories.
    }
    \label{fig:fit}
\end{figure}

The signal yield $N_{\pipi\mumu}$ is obtained from a simultaneous, extended unbinned maximum-likelihood fit to the four-track invariant-mass distributions of the signal candidates in the two trigger categories.
The invariant-mass distributions of the selected candidate events, and the result of the described mass fit are shown in \figref{fig:fit}. 
No physical candidates with an invariant mass lower than $490~\mevcc$ can exist due to conservation of energy.
To account for this, the invariant-mass distribution of the background 
follows the distribution of the available momentum in a two-body \decay{\KS}{(\pipi)(\mumu)} decay. This distribution is modelled by a function with only one free parameter which dictates the kinematic threshold of the candidates,
\begin{equation}
    \label{eq:bkgpdf}
    Q(m, A) = \begin{cases}
    \sqrt{\Delta}&~\text{if}~\Delta > 0\\
    0 &~\text{if}~\Delta < 0
    \end{cases}~,
\end{equation}
with
\begin{equation}
    \Delta = \frac{ [(m - A)^2 - (2m_\muon + 2m_\pion)^2][(m - A)^2 - (2m_\muon - 2m_\pion)^2] }{4(m - A)^2}\,,
\end{equation}
where $m$ is the four-track invariant mass $m(\pipi\mumu)$,
and $A$ is its  minimum value, which can exceed the sum of the masses of the two pions and two muons due to the requirements on the minimum angle between the decay products. Specifically, $Q$ is the centre-of-mass momentum in a two-body decay, introducing an offset, $A$, on the mass of the origin particle, $m$, that models the kinematic threshold for the detector.
The $A$ parameter is extracted from the fit.
The background model is validated on background-enriched data obtained by loosening the BDT selection requirement. 
The \Kppmm signal invariant-mass shape is parametrised by the sum of two Crystal Ball (CB) functions~\cite{Skwarnicki:1986xj}, with the same peak position, and power-law tails on opposite sides of the distribution. 
The parameter values of the signal distribution
are obtained from simulated events with corrections 
obtained from a fit \Kspipi decays in data, following Ref.~\cite{LHCb-PAPER-2019-038}.

\begin{figure}[tb]
    \centering
    \includegraphics[width=0.7\linewidth]{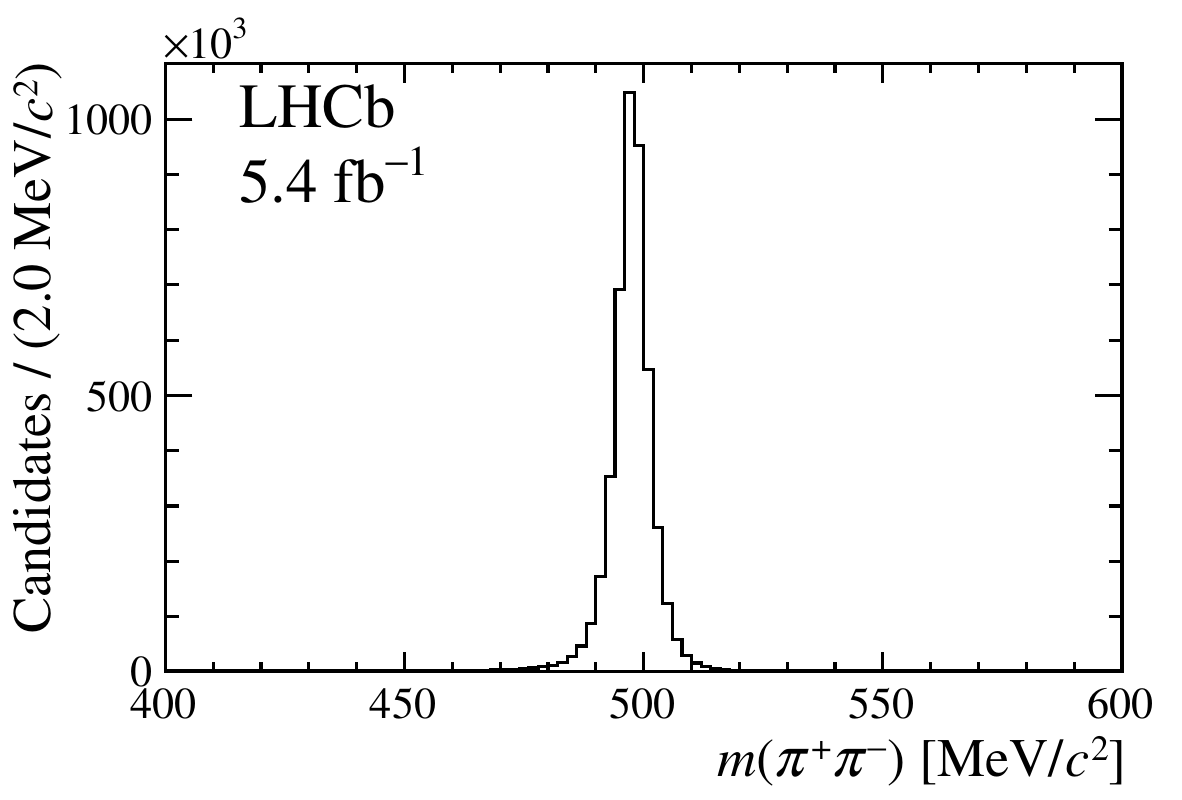}
    \caption{Invariant-mass distribution of the selected \Kspipi candidates.}
    \label{fig:mpipi}
\end{figure}

The \Kspipi candidates, selected by combining two well-identified pions with opposite charge originating from a common vertex, are further restricted to specific regions of the Armenteros--Podolanski plane~\cite{armenteros} in order to suppress contributions from \Lambdappi and \Lambdabppi decays, as seen in Fig.~\ref{fig:AP}. The details of the selection used can be found in Appendix~\ref{sec:ap_req}.
\begin{figure}[tb]
    \centering
    \includegraphics[width=0.7\linewidth]{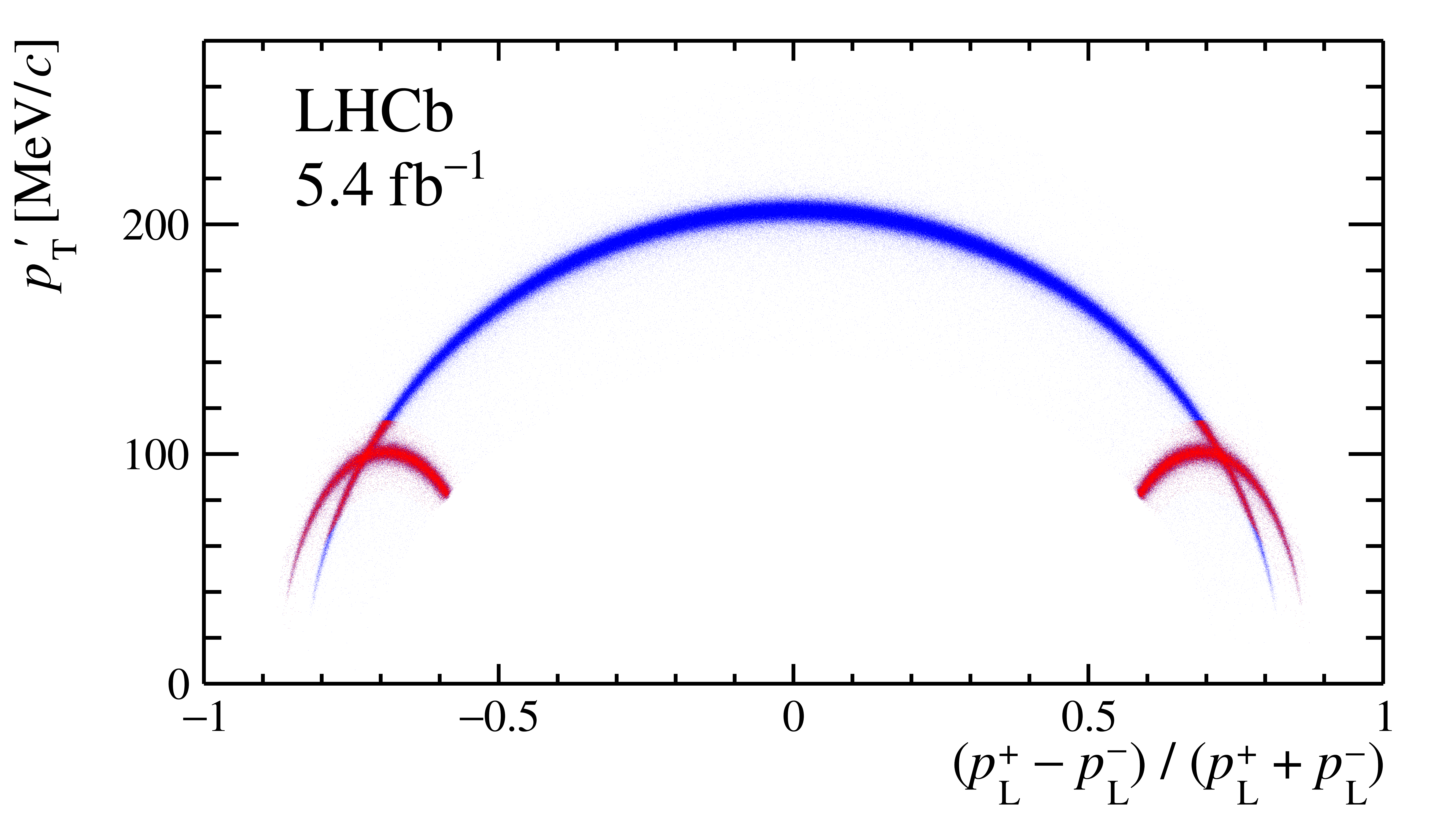}
    \caption{Distribution of selected \Kspipi candidates in data in the Armenteros--Podolanski plane. The variables of the plane are the longitudinal momentum asymmetry between the positively ($p_\text{L}^+$) and negatively ($p_\text{L}^-$) charged pions,
    and the transverse momentum of the pions with respect to the \KS flight direction, $p_\text{T}'$. The blue region, containing $\Kspipi$ decays, is accepted, while the red regions, containing background from \Lambdappi and \Lambdabppi decays, are rejected.}
    \label{fig:AP}
\end{figure}
The $N_{\pipi}$ value is determined as the number of candidates in the mass range $400<m(\pipi)<600\mevcc$. Due to the negligible background contribution after the selection in the \Kspipi sample, as illustrated by Fig.~\ref{fig:mpipi}, no fit to the invariant-mass distribution is performed to determine the number of \Kspipi decays. 


The efficiencies are assumed to factorise according to
\begin{equation}
\varepsilon = \varepsilon^{\text{reco}}\varepsilon^{\text{sel$/$reco}}\varepsilon^{\text{trig$/$sel}}.
\end{equation}
Here $\varepsilon^{\text{reco}}$ is the reconstruction efficiency,
$\varepsilon^{\text{sel/reco}}$ is the selection efficiency for the reconstructed signal candidates, and $\varepsilon^{\text{trig/sel}}$ 
is the trigger efficiency for selected candidates.
The $\varepsilon^{\text{reco}}$ efficiency includes track and vertex reconstruction effects and is estimated from simulation to be $\sim0.2\%$ for \Kppmm decays and $\sim1.5\%$ for \Kspipi decays. The $\varepsilon^{\text{sel/reco}}$ efficiency, for \Kppmm decays, includes the BDT and PID requirements and is estimated to be around $7\%$. 
The equivalent efficiency for \Kspipi decays  includes the cut on the Armenteros--Podolanski plane, and it is approximately $46\%$.
Lastly, the $\varepsilon^{\text{trig/sel}}$ 
 efficiency is determined to be around $1\%$ for \Kppmm decays. The unbiased trigger selection used for the \Kspipi mode is too loose to record every selected event. The acceptance rate is artificially reduced by considering only one event in every $10^{-7}$ collision. This is considered as trigger efficiency for the \Kspipi mode.
The efficiencies are calculated using simulation samples with corrections derived from \Kspipi decays in data. 
The tracking efficiency is obtained using dedicated control samples of  $J/\psi\rightarrow\mumu$ and \Kspipi decays using a tag-and-probe method~\cite{LHCb-DP-2013-002}. Similarly, PID efficiencies are obtained using calibration samples of low-momentum muons from $\jpsi\rightarrow\mumu$ decays collected concurrently with the data used for the analysis. 
The relatively low selection efficiency for the \Kppmm channel is primarily driven by the limited phase space available; the small mass difference between the \KS and the four final-state particles results in low-\pt final-state products and small opening angles. This makes it difficult to distinguish signal tracks from the background produced in \proton\proton collisions without a significant loss in efficiency.
The trigger efficiencies obtained in simulation are validated with 
\decay{\Kz}{\pipm\mump\nuornubar_{\hspace{-6pt}\muon}} 
decays, by comparing the trigger efficiency in simulation and data, as described in Ref.~\cite{LHCb-PAPER-2019-038}. 
The simulated samples used to compute the efficiencies are generated assuming the decay amplitude is uniform over the phase space of the \KS particle without accounting for the presence of any intermediate resonances. The obtained efficiencies are verified to be independent of such effects.
The efficiency for the \Klppmm decay is derived using the same procedure applied to \Kppmm simulated samples, weighted to reproduce the \KL lifetime, as mentioned in Sec.~\ref{sec:lhcb}.

\section{Systematic uncertainties}
\label{sec:systematics}

\tabref{tab:SystNorm} shows the sources of systematic uncertainty on $\alpha_\text{S}$ considered in this analysis. 
The systematic uncertainties are incorporated
into the fit of the signal-candidate mass distribution
as Gaussian priors on the single event sensitivity.
For convenience, sources of systematic uncertainty are separated into those acting independently on each trigger category 
and those affecting the overall normalisation factor.
\begin{table}[b]
  \centering
  \caption{Summary of the sources of systematic uncertainty,
  as well as their relative effect on the single event sensitivity, $\alpha_\text{S}$.}
  \label{tab:SystNorm}
  \begin{tabular}{lc }
  \toprule
    Source                 & Relative effect $(\%)$                        \\
    \midrule
    Variations in data taking      & $1$                   \\
    Data/simulation differences & $9$                     \\
    \Kspipi yield    & $1$                                      \\
    PID                 & $2$    \\
    Tracking            &  $2$\\
    \lone     & $9$ (TIS), $11$ (TOS)                      \\
    \hlt & $13$     \\
  
    \midrule
    Total                  & $18$ (TIS), $19$ (TOS) \\ 
    \bottomrule
  \end{tabular}
\end{table}

  A $1\%$ systematic uncertainty on the efficiencies accounts for minor modifications of the trigger selection criteria throughout the Run~2 data-taking period. These are determined by comparing the yields of triggered   \decay{\Kz}{\pipm\mump\nuornubar_{\hspace{-6pt}\muon}} and \Kspipi decays from the trigger-unbiased data.
  To compute the systematic uncertainty associated with the corrections applied to the simulation, 
  the efficiency ratio is recomputed with and without corrections on the kaon momenta. 
  Half of the correction is taken as the systematic uncertainty and amounts to a 9\% effect on $\alpha_{\rm S(L)}$.
  The systematic uncertainty on the \Kspipi yield is evaluated 
  by comparing the number of selected \Kspipi candidates in data with the yield obtained from a fit to the \pipi invariant mass with a floating background component modelled as an exponential function.
  A 1\% systematic uncertainty contribution is found to cover any variation.
  The uncertainty on the PID efficiency corrections is propagated to the ratio of efficiencies and gives an additional 2\% systematic uncertainty.
  The differences between data and simulation in the track reconstruction efficiencies,
  determined using control channels from data, are covered by a $2\%$ systematic uncertainty.
  The \lone and \hlt efficiencies for the \mbox{\Kppmm} signal are determined using 
  \mbox{\decay{\Kz}{\pipm\mump\nuornubar_{\hspace{-6pt}\muon}}}~decays as control modes. The efficiencies are found to be compatible between data and simulation, and their difference is used as the systematic uncertainty, following Ref.~\cite{LHCb-PAPER-2019-038}. The  
  \mbox{\decay{\Kz}{\pipm\mump\nuornubar_{\hspace{-6pt}\muon}}} sample at \lhcb is composed of \KS and \KL in similar amounts, due to a compensation between the different branching fractions~\cite{KLOE:2005lau, KLOE:2005vdt,KLOE-2:2019rev}, and the different reconstruction efficiency driven by the decay-time acceptance, which produces similar decay-time distributions for reconstructed \KL and \KS decays. 
A systematic uncertainty of $1.4\%$ is added to the relative efficiency between \mbox{\Kppmm} and \mbox{\Klppmm} decays, when measuring \mbox{\BRof\Klppmm}. This systematic uncertainty covers the difference of the efficiency ratio across the data-taking period as well as uncertainties related to the limited simulation sample size. The systematic uncertainties from the signal and background invariant-mass models are negligible at the current level of precision. 
The uncertainty on the branching fraction \mbox{$\BRof\Kspipi  = (69.20 \pm 0.05) \%$} induces a $0.07\%$ uncertainty on $\alpha_{\rm S(L)}$, also considered negligible. 
\section{Results and conclusions}
\label{sec:results}

The number of trigger-efficiency corrected \Kspipi decays is found to be \mbox{$(8.79 \pm0.11) \times 10^{12}$}.
The efficiency ratios, $\varepsilon_{\Kspipi}/\varepsilon_{\Kppmm}$, are determined to be $(2.9\pm0.3)\times10^3$ and $(5.4\pm0.5)\times10^3$ for the TIS and TOS categories, respectively. Note that the trigger efficiency is considered in the \Kspipi yield instead of on the efficiency ratios.
The obtained normalisation factors are 
\mbox{$\alpha_{\text{S}}^{\text{TIS}} = (2.27\pm 0.42) \times 10^{-10}$} and $\alpha_{\text{S}}^{\text{TOS}} = (4.21\pm 0.81) \times 10^{-10}$, 
assuming that the muons and pions are uniformly distributed in phase space for \mbox{\Kppmm} decays. 
The normalisation factors, $\alpha_{\text{TIS}}$ and $\alpha_{\text{TOS}}$, are derived for the \Klppmm decay, using Eq.~\ref{eq:alpha}, and the efficiency for \Klppmm decays, computed as described in Sec.~\ref{sec:normalisation}. 

The final data is consistent with the background-only hypothesis at the level of $1.9$ standard deviations, showing no evidence for a signal.
Upper limits are set, at 90\% CL, by integrating the exponent of the negative profiled log-likelihood up to 90\% of its total area, validated using pseudoexperiments. The upper limits are determined to be:
\begin{align}
    \BRof\Kppmm &< 1.4 \times 10^{-9}, \nonumber \\
    \BRof\Klppmm &< 6.6 \times 10^{-7} \nonumber ,
\end{align}
where the limit on each decay mode assumes the absence of the other.

In summary, a search for the \Kzppmm decays has been performed using $pp$ collision data recorded at \lhcb between 2016 and 2018, corresponding to an integrated luminosity of  $5.4\invfb$. 
No evidence for signal is found, and an upper limit for each mode is reported for the first time.

\section*{Acknowledgements}
%
%
\noindent We express our gratitude to our colleagues in the CERN
accelerator departments for the excellent performance of the LHC. We
thank the technical and administrative staff at the LHCb
institutes.
We acknowledge support from CERN and from the national agencies:
ARC (Australia);
CAPES, CNPq, FAPERJ and FINEP (Brazil); 
MOST and NSFC (China); 
CNRS/IN2P3 (France); 
BMFTR, DFG and MPG (Germany); 
INFN (Italy); 
NWO (Netherlands); 
MNiSW and NCN (Poland); 
MCID/IFA (Romania); 
MICIU and AEI (Spain);
SNSF and SER (Switzerland); 
NASU (Ukraine); 
STFC (United Kingdom); 
DOE NP and NSF (USA).
We acknowledge the computing resources that are provided by ARDC (Australia), 
CBPF (Brazil),
CERN, 
IHEP and LZU (China),
IN2P3 (France), 
KIT and DESY (Germany), 
INFN (Italy), 
SURF (Netherlands),
Polish WLCG (Poland),
IFIN-HH (Romania), 
PIC (Spain), CSCS (Switzerland), 
and GridPP (United Kingdom).
We are indebted to the communities behind the multiple open-source
software packages on which we depend.
Individual groups or members have received support from
Key Research Program of Frontier Sciences of CAS, CAS PIFI, CAS CCEPP, 
Fundamental Research Funds for the Central Universities,  and Sci.\ \& Tech.\ Program of Guangzhou (China);
Minciencias (Colombia);
EPLANET, Marie Sk\l{}odowska-Curie Actions, ERC and NextGenerationEU (European Union);
A*MIDEX, ANR, IPhU and Labex P2IO, and R\'{e}gion Auvergne-Rh\^{o}ne-Alpes (France);
Alexander-von-Humboldt Foundation (Germany);
ICSC (Italy); 
Severo Ochoa and Mar\'ia de Maeztu Units of Excellence, GVA, XuntaGal, GENCAT, InTalent-Inditex and Prog.~Atracci\'on Talento CM (Spain);
SRC (Sweden);
the Leverhulme Trust, the Royal Society and UKRI (United Kingdom).

\section*{Appendices}

\appendix
\section{Armenteros--Podolanski requirement}
\label{sec:ap_req}
As described in Sec.~\ref{sec:normalisation}, the \Kspipi sample is selected using a requirement in the Armenteros--Podolanski plane. Such cut is defined as
\begin{equation*}
 \left| \left[ \left( \left(\alpha \pm \dfrac{M_p^2-M_{\pi}^2}{M_{\Lambda}^2}\right)
    \dfrac{M_{\Lambda} p_{K_S^0}} {2 p^* \sqrt{p_{K_S^0}^2+M_{\Lambda}^2}} \right)^2 + \dfrac{p_{\rm T}^2}{(p^*)^2} \right] -1 \right| > 0.3  \, ,
\end{equation*}
where, $\alpha = \frac{p_{\rm L}^+ - p_{\rm L}^-}{p_{\rm L}^+ + p_{\rm L}^-}$ is the longitudinal momentum asymmetry between the positive and negative final state tracks, $p_{\rm L}$ and $\pt$ are the longitudinal and transverse track momentum with respect to the origin particle momentum in the laboratory frame, and $p^*\equiv\left[ (M_{\Lambda}^2-M_p^2-M_{\pi}^2)^2 - 4 M_p^2 M_{\pi}^2 \right]/(4 M_{\Lambda}^2)$ is the magnitude of final state track momentum in the centre-of-mass frame, assuming a \Lz decay.


\addcontentsline{toc}{section}{References}
\bibliographystyle{LHCb}
\bibliography{main,standard,LHCb-PAPER,LHCb-CONF,LHCb-DP,LHCb-TDR}

\newpage
\centerline
{\large\bf LHCb collaboration}
\begin
{flushleft}
\small
R.~Aaij$^{38}$\lhcborcid{0000-0003-0533-1952},
A.S.W.~Abdelmotteleb$^{58}$\lhcborcid{0000-0001-7905-0542},
C.~Abellan~Beteta$^{52}$\lhcborcid{0009-0009-0869-6798},
F.~Abudin{\'e}n$^{58}$\lhcborcid{0000-0002-6737-3528},
T.~Ackernley$^{62}$\lhcborcid{0000-0002-5951-3498},
A. A. ~Adefisoye$^{70}$\lhcborcid{0000-0003-2448-1550},
B.~Adeva$^{48}$\lhcborcid{0000-0001-9756-3712},
M.~Adinolfi$^{56}$\lhcborcid{0000-0002-1326-1264},
P.~Adlarson$^{86}$\lhcborcid{0000-0001-6280-3851},
C.~Agapopoulou$^{14}$\lhcborcid{0000-0002-2368-0147}
% ...

{\footnotesize \it

% ...
$^{14}$Universit{\'e} Paris-Saclay, CNRS/IN2P3, IJCLab, Orsay, France\\
% ...
$^{38}$Nikhef National Institute for Subatomic Physics, Amsterdam, Netherlands\\
% ...
$^{48}$Instituto Galego de F{\'\i}sica de Altas Enerx{\'\i}as (IGFAE), Universidade de Santiago de Compostela, Santiago de Compostela, Spain\\
% ...
$^{52}$Physik-Institut, Universit{\"a}t Z{\"u}rich, Z{\"u}rich, Switzerland\\
% ...
$^{56}$H.H. Wills Physics Laboratory, University of Bristol, Bristol, United Kingdom\\
% ...
$^{58}$Department of Physics, University of Warwick, Coventry, United Kingdom\\
% ...
$^{61}$School of Physics and Astronomy, University of Glasgow, Glasgow, United Kingdom\\
$^{62}$Oliver Lodge Laboratory, University of Liverpool, Liverpool, United Kingdom\\
% ...
$^{70}$Syracuse University, Syracuse, NY, United States\\
% ...
$^{86}$Department of Physics and Astronomy, Uppsala University, Uppsala, Sweden, associated to $^{61}$\\
% ...
\medskip
}
\end{flushleft}

\end{document}